# Superparticles from the Initial Universe and deduction of the Fine Structure Constant and Uncertainty Principle directly from the Gravitation Theory.

## Fran De Aquino[*]

Physics Department, Maranhao State University, S.Luis/MA,Brazil.

**ABSTRACT**
In a previous work it was shown that the gravitational and inertial masses are correlated by an adimensional factor, which depends on the incident radiation upon the particle. It was also shown that there is a direct correlation between the radiation absorbed by the particle and its gravitational mass, independently of the inertial mass. This finding has fundamental consequences to Unified Field Theory and Quantum Cosmology. Only in the absence of electromagnetic radiation the mentioned factor becomes equal to *one*. On the other hand, in specific electromagnetic conditions, it can be reduced, nullified or made negative. This means that there is the possibility of the gravitational masses can be reduced, nullified and made negative by means of electromagnetic radiation. This unexpected theoretical result was recently confirmed by an experiment (gr-qc/0005107). A fundamental consequence of the mentioned correlation is that, in specific ultra-high energy conditions, the gravitational and electromagnetic fields can be described by the same Hamiltonian, i.e., in these circumstances, they are *unified* ! Such conditions can have occurred *inclusive* in the Initial Universe, before the first spontaneous breaking of symmetry. Taking as base this discovery, and starting from the *gravitational mass* of *superparticles* from the Initial Universe we show here that it is possible to deduce the *reciprocal fine structure constant* and the *uncertainty principle* directly from the Gravitation Theory(Unified Theory).

## INTRODUCTION

In a recent paper[1] we have shown that the *gravitational mass* and the *inertial mass* are correlated by an adimensional factor, which depends on the incident radiation upon the particle. It was shown that only in the absence of electromagnetic radiation this factor becomes equal to 1 and that, in specific electromagnetic conditions, it can be reduced, nullified or made negative.

The general expression of correlation between gravitational mass $m_g$ and inertial mass $m_i$, is given by

$$m_g = m_i - 2\left\{\sqrt{1+\left[\frac{U}{m_i c^2}\sqrt{\frac{\varepsilon_r \mu_r}{2}\left(\sqrt{1+(\sigma/\omega\varepsilon)^2}+1\right)}\right]^2} - 1\right\}m_i \quad (I)$$

where $U$ is the electromagnetic energy absorbed by the particle; $\varepsilon = \varepsilon_r \varepsilon_0$, $\mu = \mu_r \mu_0$ and $\sigma$ are the electromagnetic characteristics of the outside medium around the particle in which the incident radiation is propagating.

In the GUTs, the Initial Universe was simplified for just two types of fundamental particles: the *boson* and the *fermion*. However, bosons and fermions are unified in *Supergravity*: One can be transformed into another, just as quarks can be transformed into leptons in the GUTs. Thus, in the period where gravitation and electromagnetism were unified. (which would have occurred from time zero up to a critical time $t_c \cong 10^{-43}$s after Big-Bang), the Universe should have been extremely simple – with just *one* particle type ( *superparticle* or *protoparticle* ).

Starting from the *gravitational mass* of these superparticles we will

---

show that the *reciprocal fine structure constant* and the *uncertainty principle* can be deduced directly from the Gravitation Theory( Unified Theory).

## 1.CALCULATIONS

According to equation (1) the *gravitational mass* of a particle in the *free space* ($\varepsilon_r=\mu_r=1$ and $\sigma=0$) is given by:

$$m_g = m_i - 2\left\{\sqrt{1+\left[\frac{U}{m_i c^2}\right]^2} - 1\right\}m_i \quad (2)$$

In the case of thermal radiation, it is usual to relate the energy of photons to temperature, through the relation, $<h\nu> \sim kT$ where $k=1.38 \times 10^{-23}$ J/K is the Boltzmann's constant. Thus, in this case, the energy absorbed by the particle will be $U=\eta<h\nu>\sim\eta kT$, $\eta$ is a particle-dependent absorption coefficient (in general, for elementary particles, $\eta \sim 0.1$).

The temperature T of the Universe in the $10^{-43}$s < t < $10^{-23}$s period can be calculated by means of the well-known expression[2]

$$T \sim 10^{22}(t/10^{-23})^{-1/2}. \quad (3)$$

This means that in the period where gravitation and electromagnetism were unified (which would have occurred from time zero up to a critical time $t_c \cong 10^{-43}$s after Big-Bang ) the temperature was T >> $10^{32}$K .

According to the Hawking's[3] prediction, collapsed objects cannot have Schwarzchild's radius less than the particle's *wave length*. This correspond to *inertial mass* $m_i \sim 10^{-8}$kg. On the other hand, the highest temperature T >> $10^{32}$K indicates that $\eta kT/c^2 >> 10^{-8}$kg. Thus, we can assume that $U = \eta kT >> m_{sp}c^2$. ($m_{sp}$ is the *superparticle's inertial mass*.).

For $U=\eta kT >> m_{sp}c^2$ the equation (2) reduces to:

$$m_g \cong -2U/c^2 = -2\eta kT/c^2 >> 10^{-8}\text{kg} \quad (3)$$

which is the *gravitational mass* of the *superparticles*.

It follows that the gravitational forces between two *superparticles* is given by:

$$\vec{F}_{12} = -\vec{F}_{21} = G\frac{m_g m_g'}{r^2}\hat{\mu}_{21}$$

$$= \left[\left(\frac{4G}{c^5\hbar}\right)(\eta\kappa T)^2\right]\frac{\hbar c}{r^2}\hat{\mu}_{21} \quad (4)$$

Due to the gravitational and electromagnetic interactions were themselves *unified* in that period, we can write,

$$\left|\vec{F}_{12}\right| = \left|-\vec{F}_{21}\right| = \left|\left[\left(\frac{4G}{c^5\hbar}\right)(\eta\kappa T)^2\right]\frac{\hbar c}{r^2}\hat{\mu}_{21}\right|$$

$$= \frac{e^2}{4\pi\varepsilon_0 r^2} \quad (5)$$

From equation above we can write

$$\left[\left(\frac{4G}{c^5\hbar}\right)(\eta\kappa T)^2 \hbar c\right] = \frac{e^2}{4\pi\varepsilon_0} \quad (6)$$

If now we make in the above equation

$$\left(\frac{4G}{c^5\hbar}\right)(\eta\kappa T)^2 = \alpha \quad (7)$$

it becomes

$$\alpha = \frac{e^2}{4\pi\varepsilon_0 \hbar c} = \frac{1}{137} \quad (8)$$

which is the well-known *reciprocal fine structure constant* .

For T~$10^{32}$K the equation(7) tell us that

$$\alpha = \left(\frac{4G}{c^5\hbar}\right)(\eta\kappa T)^2 \cong \frac{1}{100} \quad (9)$$

This value has the same order of magnitude that the exact value(1/137) of the *reciprocal fine structure constant*.

From equation (4) we can write:

$$\left(G\frac{m_g m_g'}{\alpha c \vec{r}}\right)\vec{r} = \hbar \quad (10)$$

The term between parenthesis has the same dimensions that the *linear momentum* $\vec{p}$. Thus Eq.(10) gives us

$$\vec{p}\cdot\vec{r} = \hbar \quad (11)$$

A component of the momentum of a particle cannot be precisely specified without loss of all knowledge of the corresponding component of its position at that time ,i.e., a particle cannot precisely localized in a particular direction without loss of all knowledge of its momentum component in that direction . This means that in intermediate cases the product of the uncertainties of the simultaneously measurable values of corresponding position and momentum components is *at least of the order of magnitude* of $\hbar$ ,i.e.,

$$\Delta p \cdot \Delta r \geq \hbar \quad (12)$$

This relation, *directly obtained here from the Unified Theory*, is the well-known relation of the *Uncertainty Principle* for position and momentum.

Finally, it can be shown by standard mathematical methods that

$$\Delta k \cdot \Delta x \geq 1 \quad (13)$$

where $\Delta k$ is the approximate spread in *propagation number* $k = 2\pi/\lambda$.

When we combine the relation(12), with the relation(13), we obtain the equations $\vec{p} = \hbar\vec{k}$ and $E = hf$, that are the so-called *De Broglie-Einstein relations.* Therefore the Quantum Mechanics can be deduced from the New Unified Theory.

## 2.CONCLUSION

In the present paper we have shown that it is possible to deduce the *reciprocal fine structure constant and the uncertainty principle* directly from the Unified Theory .

These are theoretical findings that validate the discovery of the correlation between Gravitation and Electromagnetism and confirm the Unified Theory, both carried out by the author.[1]

## 3.REFERENCES